\documentstyle[preprint,revtex]{aps}
\begin{document}

%%%%%%letter d in Ruder%%%%%%%%%%%%%%%%%%%%%%%%%
\setbox16= \hbox{der \kern -17.2pt \raise %
 3.8pt \hbox{-} \vrule width -2.9pt depth 0pt }
\def\dj{\copy16}
%%%%%%%%%%%%%%%%%%%%%%%%%%%%%%%%%%%%%%%%%%%%%%%%

\renewcommand{\thepage}{\roman{page}}
\setcounter{page}{1}

\draft

\today \hfill LBL-32345

\hfill UCB-PTH-92/18

\begin{title}
Flavor changing interactions mediated by scalars at the weak scale
\end{title}

\author{Aram Antaramian, Lawrence J. Hall and
Andrija Ra\v{s}in \( \!\! ^{(a)} \) }

\begin{instit}
Department of Physics \\
University of California, Berkeley \\
and \\
Lawrence Berkeley Laboratory \\
1 Cyclotron Road, Berkeley, CA 94720
\end{instit}

%%%%%%%%%%%%%%%%
\begin{abstract}
%%%%%%%%%%%%%%%%
The quark and lepton mass matrices possess approximate flavor symmetries.
Several results follow if the interactions of new scalars possess these
approximate symmetries. Present experimental bounds allow these exotic
scalars to have a weak scale mass. The Glashow-Weinberg criterion is rendered
unnecessary. Finally, rare leptonic B meson decays provide powerful probes of
these scalars, especially if they are leptoquarks.
\end{abstract}


\newpage

\begin{center}
{\bf Disclaimer}
\end{center}

\vskip .2in

\begin{scriptsize}
\begin{quotation}
This document was prepared as an account of work sponsored by the United
States Government.  Neither the United States Government nor any agency
thereof, nor The Regents of the University of California, nor any of their
employees, makes any warranty, express or implied, or assumes any legal
liability or responsibility for the accuracy, completeness, or usefulness
of any information, apparatus, product, or process disclosed, or represents
that its use would not infringe privately owned rights.  Reference herein
to any specific commercial products process, or service by its trade name,
trademark, manufacturer, or otherwise, does not necessarily constitute or
imply its endorsement, recommendation, or favoring by the United States
Government or any agency thereof, or The Regents of the University of
California.  The views and opinions of authors expressed herein do not
necessarily state or reflect those of the United States Government or any
agency thereof of The Regents of the University of California and shall
not be used for advertising or product endorsement purposes.
\end{quotation}
\end{scriptsize}

\vskip 2in

\begin{center}
\begin{small}
{\it Lawrence Berkeley Laboratory is an equal opportunity employer.}
\end{small}
\end{center}

\newpage

\newpage
\renewcommand{\thepage}{\arabic{page}}
\setcounter{page}{1}

\widetext

%%%%%%%%%%%%%%%%%%%%%%
\section{Introduction}
%%%%%%%%%%%%%%%%%%%%%%
As more and more tests of the standard model confirm its predictions to ever
higher accuracy, it becomes tempting to believe that new physics, especially
if it involves flavor changing neutral currents, can only
occur at energy scales
very much larger than the weak scale. For example, $\Delta S = 2$ four fermion
operators with coefficients $1/\Lambda^2$ give a $K_L - K_S$ mass difference
$\Delta m_K/m_K \simeq (f_K/\Lambda)^2$ implying that $\Lambda \geq 1000$ TeV.
The purpose of this paper is to show that it is perfectly natural for
physics involving new heavy scalars to occur at scales as low as the weak
scale, 250 GeV, and to show that rare leptonic B meson decays will provide an
excellent probe of this new physics.

In this paper we introduce a specific form for the way that approximate flavor
symmetries act on quarks and leptons. We then use this as a guide to infer the
expected size of couplings between the known fermions and hypothetical,
heavy scalar particles. The scalar mass M is then the
only unknown parameter in the coefficient of the
four fermion interactions induced by the exchange of this scalar.
We derive the experimental limits on M from a variety of rare
processes. The most powerful of these limits are of order the weak scale,
giving hope to the possiblity that we may discover physics at the weak
scale to be much richer than in the minimal standard model.
There are two important advantages of our general approach.
The scalar mass limits depend only on symmetry arguments and not on any
specific model. Secondly, we can identify the most promising processes for
discovering new physics in the next few years.
In particular we find that rare leptonic B decays are a very
powerful probe of these new scalar interactions. For the case of leptoquarks,
these B decays will probe masses far above the present experimental limits.

An important application of our results is to flavor changing effects
in models with many Higgs doublets\cite{glas77}.
 We find that
the approximate flavor symmetries, which we already know must be a part of
any successful model
of particle physics, are sufficient to
make it natural to have any
number of Higgs doublets coupling to up and down type quarks. In other words it
is completely unnecessary to introduce discrete symmetries which act on Higgs
doublets, as is so frequently done.

%%%%%%%%%%%%%%%%%%%%%%%%%%%%%%%%%%%%%%
\section{Approximate flavor symmetries}
%%%%%%%%%%%%%%%%%%%%%%%%%%%%%%%%%%%%%%

In the standard model, the gauge interactions of the fermions:
\begin{equation}
{\cal L}_0 = i \bar {Q}\!\!\not\!\!D Q + i \bar {U}\!\!\not\!\!D U +
      i \bar {D}\!\!\not\!\!D D + i \bar {L}\!\!\not\!\!D L +
      i \bar {E}\!\!\not\!\!D E
\end{equation}
have a global symmetry
$ U(3)_Q \times U(3)_U \times U(3)_D \times U(3)_L \times U(3)_E $ ,
where $Q_i$ and $L_i$ are $SU(2)$ doublet quarks and leptons while $U_i$ ,
$D_i$ and $E_i$ are $SU(2)$ singlets and $i = 1,2,3$ is a generation label.

The Yukawa couplings:

\begin{equation}
{\cal L} = {\cal L}_0 +
           ( \lambda^U_{ij} {\bar Q}_i U_j \frac {\tilde{H}} {\sqrt{2}} +
                \lambda^D_{ij} {\bar Q}_i D_j \frac {H} {\sqrt{2}}+
                \lambda^E_{ij} {\bar L}_i E_j   \frac {H} {\sqrt{2}} + h.c.)
\end{equation}
break
the symmetry by varying degrees down to $ U(1)_e \times U(1)_\mu
\times U(1)_\tau \times U(1)_B $.
We parametrize the approximate flavor symmetries by a set of small parameters
$ \{ \epsilon \} $, one for each of the above chiral fermion fields,
which describes the breaking
of phase rotation invariance on each fermion. Thus $\lambda^U_{ij}$ is
suppressed by both $\epsilon_{Q_i}$ and $\epsilon_{U_j}$ . The idea is
that the pattern of fermion masses and mixing angles can be described
by the set $ \{ \epsilon \} $. However, this is not a precise numerical theory
for fermion masses; equations of the form
$\lambda^U_{ij} \approx \epsilon_{Q_i} \epsilon_{U_j}$ are only meant to be
order of magnitude relations.

The Yukawa matrices $\lambda^U , \lambda^D , \lambda^L $ contain a great
deal of information about the form of the breaking of flavor symmetry.
Unfortunately, we cannot reconstruct these matrices from the information which
can be obtained from experiments, namely from the fermion masses (i.e. the
Yukawa matrix eigenvalues) and the Kobayashi-Maskawa (KM) matrix.
This information is
insufficient to derive
the form of the approximate flavor symmetries which the underlying theory
must have. Nevertheless it provides a strong guideline for giving a simple
predictive ansatz for the symmetry breaking paramaters.

The lightness of the up quark tells us that flavor symmetries strongly
suppress
the $\bar{Q}_1 U_1$ operator. However, the mass eigenvalue does not allow us to
infer whether this is because the approximate flavor symmetry is acting only on
$Q_1$, only on $U_1$, or on both. However, we need to know
whether the coefficient of a
scalar coupling to $Q_1 X$ (where $X$ is any fermion other than $U_1$) is
strongly suppressed because the up quark is very light.

We now argue that the approximate
symmetries act both on left- and right-handed fields:

\begin{itemize}
\item The flavor symmetries do not act just on the right-handed fields
      because otherwise
$U_3 \approx t_R$ couples to $(\alpha Q^u_1 +\beta Q^u_2 +\gamma Q^u_3) = t_L$
and
$D_3 \approx b_R$ couples to $(\alpha' Q^u_1 +\beta' Q^u_2 +\gamma'
Q^u_3) = b_L$
where $\alpha , \beta , \gamma , \alpha' , \beta' , \gamma' $
are arbitrary mixing angles of order one, so that
$t_L$ and $b_L$
would have no reason to be very nearly in the same
$SU(2)_L$ doublet.
\item The flavor symmetries do not act just on the left-handed fields
because in this case the approximate flavor symmetries make no distinction
between $\lambda^U$ and $\lambda^D$. A large $ m_t / m_b $ ratio could be due
to a large ratio of vevs $ v_2 / v_1 $ in a two Higgs theory, but this
would lead to an unacceptably large $ m_u / m_d $ ratio. In addition
the KM angles are given by linear mass relations such as
$\theta_c \approx m_d/m_s$ rather than the more successful
square root form  $\theta_c \approx \sqrt{m_d/m_s}$
\end{itemize}

Therefore we conclude that {\it the underlying theory must have approximate
flavor symmetries that act on both left- and right-handed fields} .

The approximate flavor symmetries and the associated set of small symmetry
breaking parameters $\{ \epsilon \}$ are defined on flavor eigenstates. In
practice it is much more useful to know what suppression factors are induced
on the mass eigenstates. Consider the up-type quarks. Assume that
$\epsilon_{Q_i} \ll \epsilon_{Q_j}$ and
$\epsilon_{U_i} \ll \epsilon_{U_j}$ for $ i < j$,
as suggested by
$ m_{U_i} \ll m_{U_j}$ . Then the mass matrix is diagonalized by unitary
rotations on the $Q_i$ by a matrix with elements
$ | V_{ij} | \approx \epsilon_{Q_i}/\epsilon_{Q_j} $
($i<j$) and on the $U_i$ by a matrix with elements
$ | V^R_{ij} | \approx \epsilon_{U_i}/\epsilon_{U_j} $
($i<j$). Relations between flavor and mass eigenstates ($Q_i'$) are
of the form:
$Q_1' = Q_1 + O(\epsilon_{Q_1}/\epsilon_{Q_2}) Q_2
 +  O(\epsilon_{Q_1}/\epsilon_{Q_3}) Q_3 $ .
This shows the important result that the flavor breaking parameters
$\{ \epsilon \}$  apply to mass eigenstates as well as to flavor eigenstates.
For example, the three flavor eigenstate contributions to $Q_1'$ all carry
the same approximate flavor symmetry suppression factor of $\epsilon_{Q_1}$.

The actual structure of the approximate low energy flavor symmetries
is likely to involve many parameters: the fermion masses and mixing angles
have very few obvious regularities. A simple predictive ansatz is shown in
table \ref{param}. It involves both
left- and right-handed fermions and is predictive because it only
involves quark and lepton masses.

The rationale behind our choice is as follows. For the leptons the flavor
symmetries on $L_i$ and $E_i$ are equally responsible for suppressing the
Yukawa couplings.
For quarks, we have again tried to have both left- and right- handed flavor
symmetries equally responsible for suppressing Yukawa couplings. However
since $Q_i$ appears in both up and down mass operators, we have taken
the symmetry breaking parameter $\epsilon_{Q_i}$ to
be the geometric mean of that expected
from $m_{U_i}$ and that expected from $m_{D_i}$.
Note that we have allowed for a two Higgs doublet model. With only one Higgs
boson $v_1 = v_2 = 250GeV$.

The ansatz gives reasonable values for the CKM mixing angles.
$V_{ij} \approx \epsilon_{Q_i}/\epsilon_{Q_j} \approx
(m_{U_i}m_{D_i}/m_{U_j}m_{D_j})^{\frac {1} {4}} , i < j$ ,
which is correct at the factor of 2 level.

One must keep in mind that the ansatz, despite
its simplicity, is hardly unique.
A more complicated ansatz might use the KM matrix as input as well
as the fermion masses. However this extra complexity is not warranted since our
ansatz is quite consistent with the KM matrix.
We use the ansatz only to
{\it estimate} the magnitudes of unknown Yukawa couplings.

%%%%%%%%%%%%%%%%%%%%%%%%%%%%%%%%%%
\section{Experimental consequences}
%%%%%%%%%%%%%%%%%%%%%%%%%%%%%%%%%%

We use our ansatz to estimate the size of the Yukawa couplings and then
the corresponding rates for various processes induced by the effective
four fermion couplings. In table \ref{exper} we list the limits on the
exchanged
scalar mass \cite{foot01} obtained from a variety of experiments .
For now we assume the
scalar exchange does induce each process and that the flavor symmetry acts only
on fermions.
Once again, we obtained these numbers using our ansatz to estimate the Yukawa
couplings, so we expect the values to be reliable up to a factor of
perhaps 2 or 3.

The factor $\kappa$ represents the ratio of the matrix element of the
new four fermion operator relative to its vacuum insertion value.
In the radiative $\mu$
decay, the $\tau$-lepton contribution dominates in the loop because it has
the largest Yukawa couplings.

First, we can see that these limits are nowhere near as strong as those
for vector exchange \cite{cahn80}. Flavor nonconserving
theories at the weak scale are not ruled out at all.
Secondly, if the uncertainty factors of 2 or 3 go the right way, it is
possible that the rare leptonic $B_s$ decay will be the first place to
discover this
new physics, considering that branching ratios $10^{-7} - 10^{-8}$ will
be obtained in the near future \cite{cdfp90}.
The branching ratio prediction for $B_s \to \mu^+ \mu^-$ is about $10^{-9}$
in the standard model, and in two Higgs doublet models with discrete
symmetries \cite{sava91}.

There are cases where the scalar cannot induce all the processes considered,
as in leptoquark models. The tree level exchange of leptoquarks generates
four fermion operators which contain two quarks and two leptons. The limits
from $K \bar{K}$ and $B \bar{B}$ mixing are therefore removed. In this case
our results are particularly important: the rare leptonic B decay modes
provide the most stringent test of models with scalar leptoquarks.

We discuss briefly the case in which approximate flavor symmetries act on
the exchanged scalar too. Such is the case in R-parity violating
supersymmetric models \cite{hall90} where the exchanged scalar is a
slepton or a squark which carries the same approximate flavor symmetry
as its fermion partners.
Then in table \ref{exper} all mass limits are lowered by an additional
symmetry breaking
factor $\epsilon_a$, where the approximate symmetry of type $"a"$ is
carried by the scalar.
It is even less likely that such theories could have been excluded.

%%%%%%%%%%%%%%%%%%%%%%%%%%%%%%%%%%%%%%%%%%%%%%%%%%%%%%%%%%%%%%%%%%%%
\section{The Glashow-Weinberg criterion for multihiggs models}
%%%%%%%%%%%%%%%%%%%%%%%%%%%%%%%%%%%%%%%%%%%%%%%%%%%%%%%%%%%%%%%%%%%%

In this section we apply our results to the case of the minimal standard
model extended only by the addition of an arbitrary number of Higgs doublets.

To avoid problems with large flavor-changing neutral currents, Glashow and
Weinberg \cite{glas77} argued that only one Higgs doublet could
couple to up-type quarks and
only one Higgs to down-type quarks. However this naturality constraint, known
as the Glashow-Weinberg criterion, was based on an unusual definition of what
is "natural". For them the avoidance of flavor-changing neutral currents was
natural in a model only if it occured for all values of the coupling constants
of that model. For us a model will be natural provided the smallness of any
coupling is guaranteed by approximate symmetries \cite{hoof80}, and we find
that this implies the Glashow-Weinberg criterion is not
necessary \cite{chen87}.

In a model with many scalar doublets it is convenient to work in a basis where
only one doublet, the Higgs doublet, acquires a vev, and all the others are
massive scalar particles which play no role in the Higgs mechanism. The
couplings of the Higgs meson are flavor-diagonal at tree level, but in
general the couplings of the other doublets are not. The limits on the mass M
of these extra scalars are given in table \ref{exper}. This shows that
{\it approximate flavor symmetries are sufficient to allow extra scalar
doublets with masses in the 100s of GeV range, there is no need for additional
symmetries to act on the scalar fields.}

One reason why this is important is that the vast majority of phenomenology
on the multihiggs models has been done assuming symmetries which force only one
Higgs to couple to up-type and one to down-type quarks. We conclude that
there is no good reason for accepting the predictions of these analyses,
except in the case of supersymmetric models.

\nonum
%%%%%%%%%%%%%%%%%%%%%
\section{conclusions}
%%%%%%%%%%%%%%%%%%%%%

In this paper we have introduced a simple ansatz for the approximate
flavor symmetries as shown in table \ref{param}. It reproduces the
KM matrix elements at the factor of 2 level. If the interactions of additional
scalars respect these approximate symmetries, then the mass limits
on the scalars from various experiments are shown in table \ref{exper} .
{}From this viewpoint, new flavor changing physics at the weak scale is not
excluded, and is natural. In particular extra Higgs doublets can couple
to both up and down type quarks; there is no need to impose additional
discrete symmetries on the scalars. We find that rare
leptonic B decay modes, such as
$B^0_s \to \mu^+ \mu^-$, could uncover this new scalar-mediated physics in the
coming years.

\newpage

We would like to thank Marjorie Shapiro for useful discussions.
LJH acknowledges partial support from the NSF Presidential Young
Investigator Program.
This work was supported in part by the Director, Office of
Energy Research, Office of High Energy and Nuclear Physics, Division of
High Energy Physics of the U.S. Department of Energy under Contract
DE-AC03-76SF00098 and in part by the National Science Foundation under
grant PHY90-21139.

\bigskip

\( ^{(a)} \) On leave of absence from the
Ru\dj \hspace{.1in} Bo\v{s}kovi\'{c} Institute, Zagreb, Croatia.

\mediumtext
\begin{table}
\caption{The ansatz for flavor symmetry breaking paramaters associated to the
chiral fermion fields.
$\eta_i = \sqrt{\frac {m_{U_i}} {v_2}}$ and
$\xi_i =\sqrt{\frac {m_{D_i}} {v_1}}$. }
\begin{tabular}{cccccc}

FIELD            & FLAVOR SYMMETRY BREAKING PARAMETER   \\
\tableline
$Q_i$      & $\sqrt{\eta_i \xi_i}$  \\
\\
$U_i$      & $\sqrt{\frac {\eta_i} {\xi_i}} \eta_i$  \\
\\
$D_i$      & $\sqrt{\frac {\xi_i} {\eta_i}} \xi_i$  \\
\\
$ L_i,E_i$     &    $\sqrt{\frac {m_{E_i}} {v_1}}$ \\
\end{tabular}
\label{param}
\end{table}

\newpage

\mediumtext
\begin{table}
\caption{Experimental lower limits on the exchanged scalar masses}
\begin{tabular}{cccccc}

process            & $M/GeV (250GeV/v_1)$  \\
\tableline
$\mu \to 3e$      & $ 1 $  \\
$\mu \to e \gamma$     & $ 4 $  \\
$\mu N \to e N$      & $ 10 $  \\
$K^0_L \to \mu^{\pm} e^{\mp}$     &    $ 20 $ \\
$B_d \to \tau^+ \tau^-$      &
$ 20 \; \; (\frac {10^{-4}} {B.R.} )^{\frac {1} {4}}$  \\
$B_s \to \mu^+ \mu^-$      & $
70 \; \; (\frac {10^{-8}} {B.R.} )^{\frac {1} {4}}$  \\
$\Delta m ( B^0_d - \bar{B}^0_d )$  & $
400 \;
\sqrt{\kappa}$  \\
$\Delta m ( K^0 - \bar{K}^0 )$   & $ 500 \;
\sqrt{\kappa}$
\end{tabular}
\label{exper}
\end{table}


\begin{thebibliography}{99}
\bibitem{glas77} S.L.Glashow and S. Weinberg, Phys. Rev. {\bf D15} 1958 (1977).
\bibitem{foot01}
For any given four fermions the exchanged scalar could induce several
four-fermion operators with different helicity structures. According to
our ansatz they could have slightly different coupling constants. Since we
do not know the relative coefficients by which these operators enter the
theory, we decided to use in our estimates the operator with the "average"
Yukawa coupling (e.q. in $\Delta m ( K^0 - \bar{K}^0 )$ we use $\bar{s}_R
d_L \bar{s}_L d_R $ rather than $ ( \bar{s}_R d_L )^2 $ or
$ ( \bar{s}_L d_R )^2 $).
\bibitem{cahn80} R.N. Cahn and H. Harari, Nucl. Phys. {\bf B176} 135 (1980);
S. Dimopoulos and J. Ellis, Nucl. Phys. {\bf B182} 505 (1981).
\bibitem{cdfp90} Proposal For An Upgraded CDF Detector, CDF/DOC/PUBLIC/1172
(1990).
\bibitem{sava91} M. Savage, Phys. Lett. {\bf B266} 135 (1991); A. Ali,
Deutsches Electronen Synchotron, Hamburg, Germany,
Preprint number: DESY-91-080.
\bibitem{hall90} For a review see L.J. Hall, Mod. Phys. Lett. {\bf A5}
467 (1990) which cites the original literature.
\bibitem{hoof80} G. t'Hooft in Recent Developments in Gauge Theories,
Carg\'{e}se Summer Institute Lectures, 1979,
eds. G. t'Hooft et al (Plenum Press, New York, 1980).
\bibitem{chen87} T.P. Cheng and M. Sher, Phys. Rev. {\bf D35}
3484 (1987) found a similar result, but only in the context
of particular models of fermion masses. Our results are model
independent and depend only on the approximate flavor symmetries.

\end{thebibliography}
\end{document}